\documentclass[twocolumn,secnumarabic,amssymb, nobibnotes, aps, prl]{revtex4}
\usepackage{epsfig}
\usepackage{amsmath,dsfont}
\usepackage{overpic}

\begin{document}
\title{\bf Comment on ``Mass and ${\bf K\Lambda}$ Coupling of the 
${\bf N^\ast}$(1535)''}
\vskip 1.5true cm
\author{A.~Sibirtsev$^{1}$}
\author{J.~Haidenbauer$^{2}$}
\author{Ulf-G. Mei{\ss}ner,$^{1,2}$}
\affiliation{ $^{1}$HISK, Universit\"at Bonn, Nu{\ss}allee 14-16, 
D-53115 Bonn, Germany}
\affiliation{ $^{2}$Institut f\"ur Kernphysik (Theorie), 
Forschungszentrum J\"ulich, D-52425 J\"ulich, Germany}


\maketitle

Recently it has been argued that the effective coupling of the 
$N^*$(1535) to $K\Lambda$ is even larger than its coupling to the
$\eta N$ channel~\cite{Liu} from an analysis of data 
on the $J/\Psi \to \bar p\Lambda K$ and $J/\Psi \to \bar pp\eta$ decays
within a resonance isobar model. Based on those couplings of $N^\ast$(1535) 
to the $\eta{N}$ and $K\Lambda$ (and the $\pi{N}$) states new properties 
of the resonance have been derived, 
namely a mass of 1400 MeV and a width of 270~MeV,
that differ radically from the standard values. 
 
As additional if not decisive argument in favor of the strong $N^\ast$(1535) 
coupling to the $K\Lambda$ channel the authors of Ref.~\cite{Liu} 
claimed that the low-energy cross section data on the reaction $pp{\to}pK^+\Lambda$ 
are compatible with a large contribution from the $N^\ast$(1535) 
resonance and that it is, in fact, needed to reproduce the experimental result. 
In their calculation based on tree-level Feynman diagrams 
the final-state interaction (FSI) between the proton and the
$\Lambda$ was neglected and it was concluded that the $N^\ast$(1535) dominates 
the reaction at low energies. 

There are several points in the procedure followed in Ref.~\cite{Liu} that
one might criticize. For example, why was only the $N^\ast$(1535) resonance 
included in the analysis of the $\bar p K\Lambda$ decay branching ratio, 
but not the $N^\ast$(1650)? The latter choice seems much more natural since
this resonance is much closer to the threshold of the 
$J/\Psi \to \bar p K\Lambda$ decay. 
However, in our comment we want to concentrate on their results for 
$pp{\to}pK^+\Lambda$. First, let us mention that their calculation 
is in contradiction with all other interpretations of the 
$pp{\to}pK^+\Lambda$ cross section where the enhancement at low energies
is due to the $\Lambda{p}$ FSI rather than the $N^\ast$(1535) 
resonance \cite{Moskal,Shyam,Sib3}. 
The result in Ref.~\cite{Liu} is also in disagreement with the experimental 
Dalitz plot~\cite{TOF} which strongly suggests that at low energies the 
$pp{\to}pK^+\Lambda$ reaction is dominated by the contribution from the 
$N^\ast$(1650) resonance.
 
\begin{figure}[t]
\vspace*{-2mm}
\centerline{\hspace*{4mm}\psfig{file=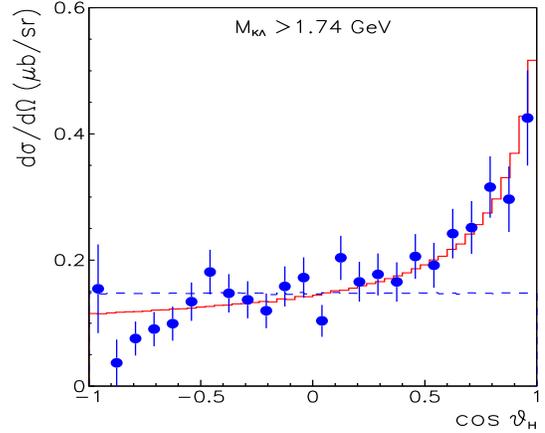,width=8.cm,height=6.5cm}}
\vspace*{-5mm}
\caption{The $\Lambda{p}$ helicity angle spectra for $K^+\Lambda$ 
masses $M_{K\Lambda}{>}$1.74 GeV for the reaction 
$pp{\to}pK^+\Lambda$ at $p_{beam}$ = 2.85 GeV/c. 
The solid histogram is our calculation including the $\Lambda{p}$ FSI, while
the dashed line is obtained without FSI but with the $N^\ast$(1535) 
parameters from Ref.~\cite{Liu}.
The data are from Ref.~\cite{TOF}. 
}
\label{fig1}
\vspace*{-5mm}
\end{figure}

In Ref.~\cite{Sib3} we proposed to study 
the role of the FSI in the $pp{\to}pK^+\Lambda$ reaction by 
analyzing differential observables, namely the Dalitz plot
distribution. By making cuts on the $K^+\Lambda$ invariant mass one 
can isolate the contribution from $N^\ast$ resonances and project 
the Dalitz plot on the $\Lambda{p}$ axis. In that case an enhancement 
at low $\Lambda{p}$ masses would manifest the presence of a FSI. It is
clear~\cite{Sib3} that $N^\ast$ resonances that couple to the
$K^+\Lambda$ system can never produce such an enhancement.

Recently the COSY-TOF Collaboration~\cite{TOF} 
measured the Dalitz plot distribution and
presented the projections on the $\Lambda{p}$ invariant mass with
different cuts on the $K^+\Lambda$ masses. The $\Lambda{p}$
spectra are shown in Ref.~\cite{TOF} in terms of the angular
distributions of the
$\Lambda$ in reference to the proton direction in the
$K^+\Lambda$ cm system, i.e. as function of the $\Lambda{p}$ helicity
angle $\theta_H$. The relation between $\theta_H$ and the
$\Lambda{p}$ invariant mass is given by Eq.~(8) of our
paper~\cite{Sib3}. Fig.~\ref{fig1}
shows experimental results for the cut on the $K^+\Lambda$
mass chosen in order to isolate the contribution from low lying 
$N^\ast$ resonances (with the convention -$\theta_H$ applied in the
experiment~\cite{TOF} so that the lowest $\Lambda{p}$ masses
corresponds to $\cos{\theta_H}$=1). 
The data indicate a strong $\theta_H$ asymmetry. 
The solid histogram is our result
obtained with inclusion of the $\Lambda{p}$ FSI. The dashed line
shows a calculation employing the $N^\ast$(1535) parameters claimed 
in Ref.~\cite{Liu} but without FSI. It is obvious that the latter 
scenario is in disagreement with the experiment. 

We consider the experimental observation of Ref.~\cite{TOF} as a direct 
evidence for the presence of the $\Lambda{p}$ FSI. Thus, the
results for $pp{\to}pK^+\Lambda$ by Liu and Zou~\cite{Liu} are rather 
questionable, and consequently also their conclusions based on the large 
$N^\ast$(1535) coupling to the $K\Lambda$ system.

\end{document}